# Reversal of the Asymmetry in a Cylindrical Coaxial Capacitively Coupled $Ar/Cl_2$ Plasma

J. Upadhyay, [1] Do Im, [1] S. Popović, [1] A.-M. Valente-Feliciano, [2] L. Phillips, [2] and L. Vušković [1]
[1]Department of Physics, Center for Accelerator Science, Old Dominion University, Norfolk, VA 23529, USA
[2]Thomas Jefferson National Accelerator Facility, Newport News, VA 23606, USA

The reduction of the asymmetry in the plasma sheath voltages of a cylindrical coaxial capacitively coupled plasma is crucial for efficient surface modification of the inner surfaces of concave three-dimensional structures, including superconducting radio frequency cavities. One critical asymmetry effect is the negative dc self-bias, formed across the inner electrode plasma sheath due to its lower surface area compared to the outer electrode. The effect on the self-bias potential with the surface enhancement by geometric modification on the inner electrode structure is studied. The shapes of the inner electrodes are chosen as cylindrical tube, large and small pitch bellows, and disc-loaded corrugated structure (DLCS). The dc self-bias measurements for all these shapes were taken at different process parameters in $Ar/Cl_2$ discharge. The reversal of the negative dc self-bias potential to become positive for a DLCS inner electrode was observed and the best etch rate is achieved due to the reduction in plasma asymmetry.

When the surface areas of two electrodes in an rf discharge are not equal, it creates an asymmetry in plasma sheath voltages, measured as a negative dc potential on a larger surface area electrode. This asymmetry has been exploited in processing of semiconductor wafers in planar plasma reactors. The dependence of the sheath voltages on the surface area of the electrodes is explained theoretically for planar geometry in Refs. [1-3] and a model for a coaxial plasma is described in Ref. [2]. The change in plasma potential and in turn the change in ion energy by applying dc voltage to an electrode in rf plasma for planar geometry is reported [4-7]. By contrast, to process the inner surface of a concave cylindrical structure serving as the grounded (outer) electrode one requires a higher potential drop in its sheath. This is achieved by changing the plasma potential of the bulk plasma.

We are using a coaxial capacitively coupled rf plasma (CCP) in $Ar/Cl_2$ mixture to develop the plasma processing of the superconducting radio frequency (SRF) cavities made of niobium (Nb). In order to modify the inner surface of the grounded outer wall, the inherent asymmetry of the coaxial CCP is compensated for and tailored in part by an external dc power supply, which has to provide an external dc current to the inner electrode in order to change its dc bias. The details regarding etch rate dependence on process parameters, etching mechanism and non-uniformity in etch rate along the gas flow direction are reported in Refs. [8, 9].

Although plasma etching of niobium on the outer cylindrical wall is possible, as shown in Refs. [8, 9], the SRF cavities present somewhat a unique challenge as they represent a cylindrical structure with variable diameter. The beam tubes in SRF cavities have a much smaller diameter compared to the maximum cavity diameter, particularly in the lower frequency elliptical SRF cavities. This poses a bottle-neck problem, where the maximum inner electrode diameter is restricted by the beam tube diameter. To overcome this problem, it is possible to increase the inner electrode surface area without increasing its diameter. Four differently structured inner electrodes were constructed and their effect on the self-bias potential was measured. Therefore, favourable processing conditions for the grounded concave electrode can be achieved by a combination of geometric and electrical corrections of the asymmetry.

The self-bias dependence on the pressure and power for rf plasma is shown in Ref. [10] for planar geometry. Its behaviour for coaxial geometry, particularly in $Ar/Cl_2$ plasma has been studied in our laboratory [11]. Here we are presenting the effect of the inner electrode surface area enhancement on the self-bias potential. The corrugated structures built on the inner electrode transform the asymmetry of the plasma by changing polarity from negative to positive self-bias potential. Its effect on the etching of the ring type samples placed on the outer electrode was remarkably beneficial. Although the effect of the size and bias of the electrode [12] and the structure of the electrode on plasma behavior [13] are studied for planar geometry processes, to the best of our knowledge, changing the polarity of the asymmetry by geometrical modification of an electrode in coaxial configuration is shown for the first time in the present work.

To understand the effect of the structure formation on the inner electrode to the asymmetry in a plasma, a coaxial type rf (13.56 MHz) plasma reactor was used with a 7.1 cm diameter and 15 cm long cylindrical vessel as the outer electrode with a variable structured inner electrode. The inner electrode is the powered electrode and the outer electrode is grounded. The matching network attached to the rf power supply not only measures the self-bias potential but has an option to attach an external dc power supply to provide a dc bias. The gases used are pure argon (Ar) and 15% chlorine ($Cl_2$) diluted in Ar. The details about the experimental setup can be found in Ref. [11].

Illustration of various structured inner electrodes are shown in Fig. 1, where (a) is 5.0 cm diameter strait tube, (b) is a standard large pitch bellows with 3.8 cm inner diameter and 4.8 cm outer diameter, and (c) is the disc–loaded corrugated structured (DLCS) electrode with a 2.5 cm inner diameter tube welded with multiple disc rings of a 5.0 cm outer diameter and 2.5 cm inner diameter. The distance between two disc rings was 3.0 mm and their thickness was 1.00 mm. Standard small pitch bellows, not shown in the Fig.1, with an outer diameter of 4.8 cm was also used in the experiment. The similar diameter of all four electrodes kept the plasma volume approximately constant and the surface area change of the inner electrode was the only parameter affecting the self-bias potential. Negative self-bias potential developed across all these four different shaped inner electrodes sheaths was measured at variable gas pressure, rf power and at two gas compositions.

The incident ion energy on the grounded wall depends on the plasma potential. The more asymmetric the plasma is, the harder it is to modify it and increase its potential to be favorable to etch outer electrode. In case of coaxial rf plasma, the etch rate is observed to decrease with the inner electrode diameter [8]. In the development of the plasma etching technology for the inner surfaces of concave 3D structures, generation of a less asymmetric plasma would be beneficial, particularly for a lower frequency SRF cavity, which has a very large diameter ellipsoid profile compared to the beam tube diameter. The ion-assisted etch rate of Nb (placed on the outer wall) for same positive dc bias is measured for various inner electrode shapes.

The self-bias potential of all four structures was measured at constant external discharge parameters, and the results are given below. The uncertainty of $Cl_2$ concentration was 2%, in rf power 3 W, in pressure 4 mTorr, in dc bias 2 V and in dc current 5 mA. In all diagrams, except in Fig. 7, the absolute value of the self-bias potential was used, although it was actually negative at the inner, powered electrode. The dc current required to make this potential zero or a positive value was recorded. The etch rate for the Nb samples placed on the outer wall for different electrodes was measured. The etch rate experiment was carried out for more than 90 minutes to avoid the fluctuation in etch rate measurements due to the lag time in starting the etching process [14].

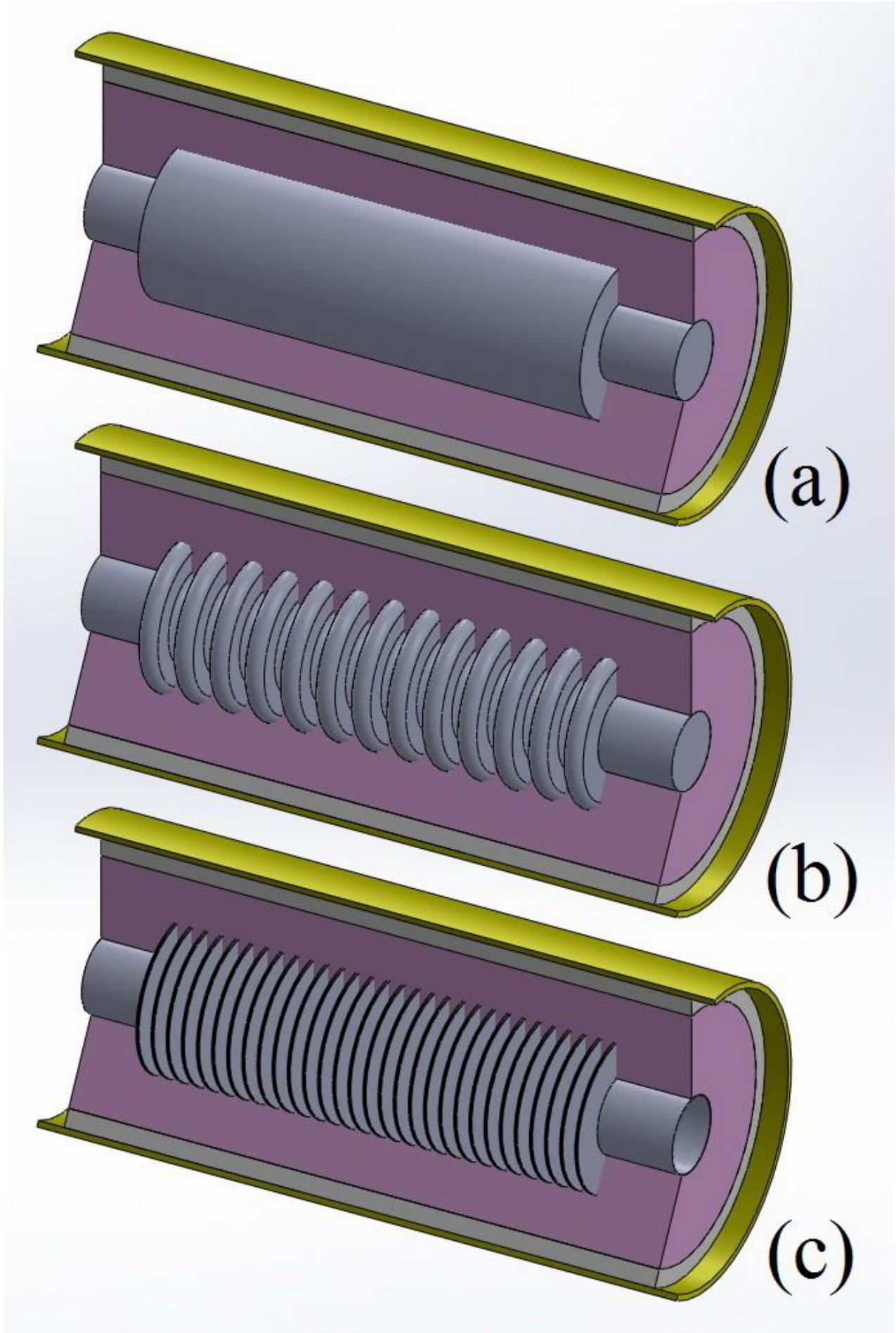

FIG. 1. Illustration of various structured inner electrodes in coaxial plasma: (a) strait tube, (b) large pitch bellows, and (c) disc-loaded corrugated structure (DLCS).

The self-bias dependence on the pressure was measured at a constant rf power of 100 W. The variation of the self-bias potential with pressure for an Ar plasma using the four different structured electrodes is shown in Fig. 2.

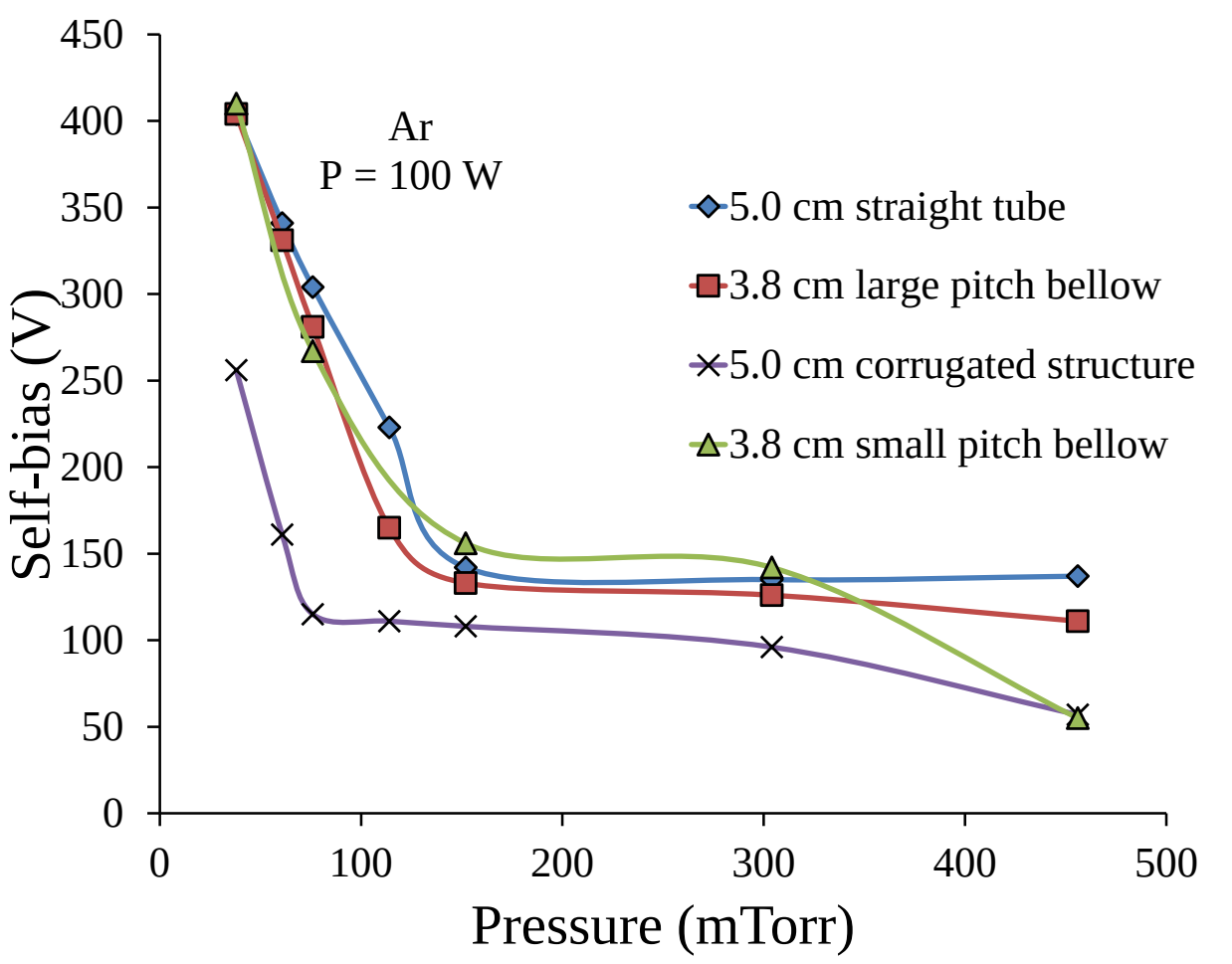

FIG. 2. Self-bias dependence on the pressure for various structured electrodes in the Ar plasma. Solid lines are visual guidelines.

Figure 2 shows that the geometrical enhancement of the inner electrode surface area reduces the bias potential compared to the straight tube electrode. It also shows that the corrugated structures are the most effective for the asymmetry reduction. Two pressure regimes described earlier in [10, 11] are clearly displayed and indicate that, while the tube electrode bias potential is constant at higher pressure, the corrugated structure and small pitched bellows show a decrease at higher pressure. The decrease in the bias potential at higher pressure could be explained by the change of the sheath thickness. As the sheath thickness decreases with pressure, it follows better the shape of the structure and the effective surface area of the inner electrode becomes larger. To explore this observation further, the rf power is varied at higher pressure for Ar plasma and the variation of the self-bias potential with the rf power for three different structures are shown in Fig. 3.

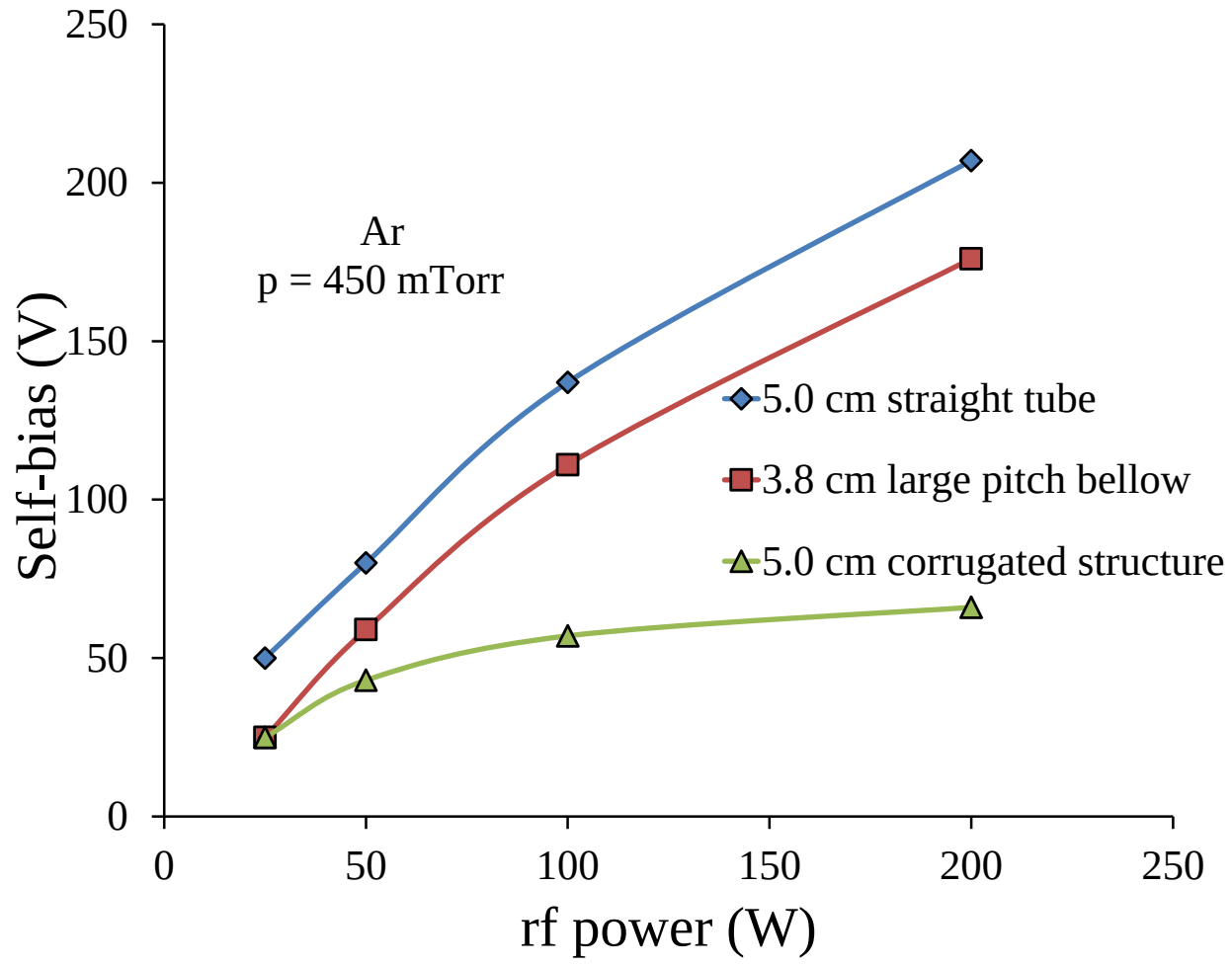

FIG. 3. Self-bias dependence in the Ar plasma on the rf power for various structured electrodes in the Ar plasma. Solid lines are visual guidelines.

Figure 3 shows that the rf power does not affect considerably the self-bias potential for the corrugated structure but has a substantial influence for the tube and large pitch bellows shaped electrode. Using the corrugated structure, not only we are increasing the effective surface area to reduce the self-bias potential but the plasma density is increased also, as described in Ref. [13], for planar geometry.

The addition of $Cl_2$ alters the electron density and the electron temperature compared to a pure Ar plasma [15-17]. The change in plasma parameters, particularly at higher pressure and rf power, leads to a decrease in the self-bias potential in the coaxial geometry [11]. In Fig. 4 is shown the pressure dependence of the self-bias potential, for rf power of 100 W, in Ar/$Cl_2$ plasma for three differently structured electrodes.

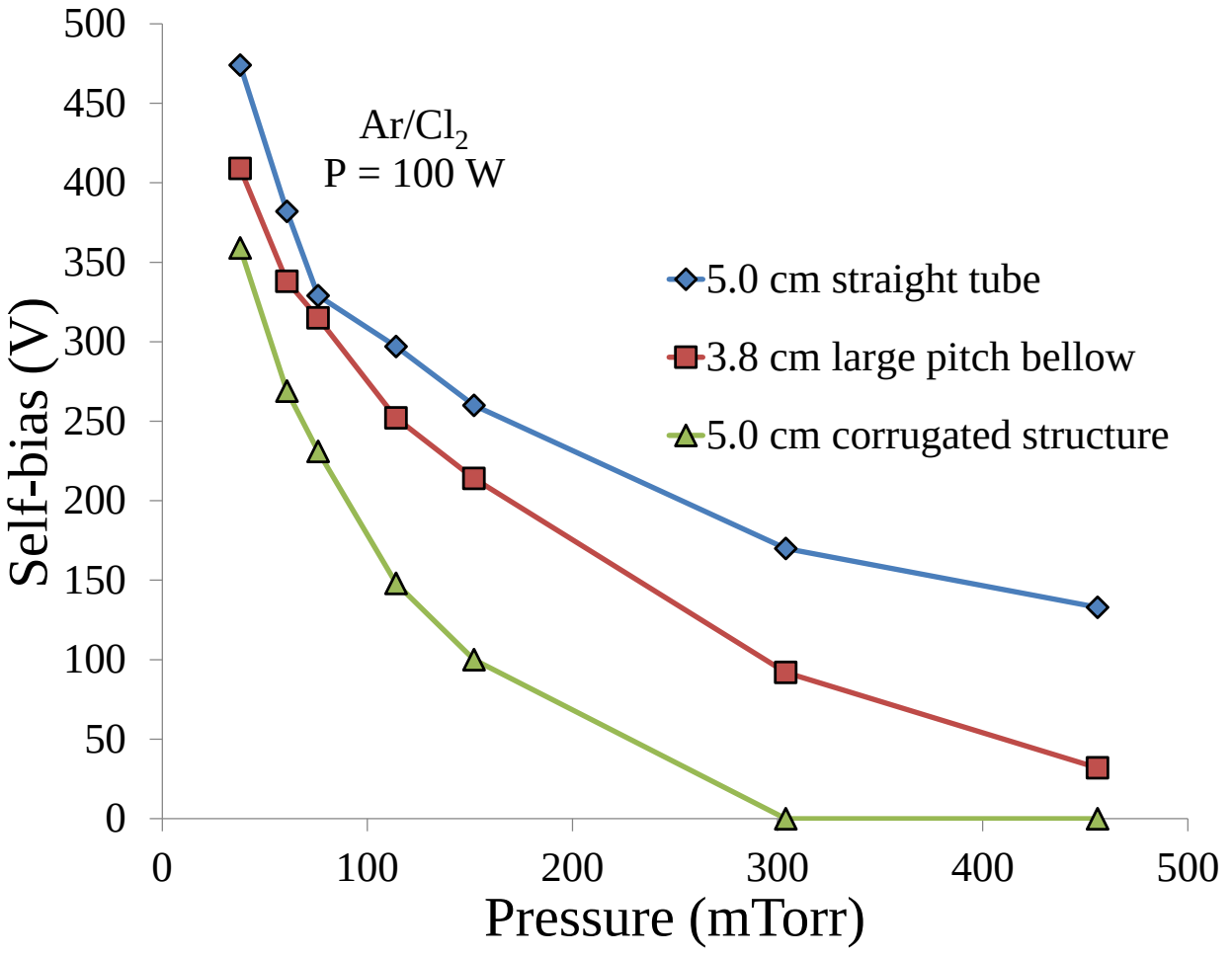

FIG. 4. Self-bias dependence on the pressure in the Ar/$Cl_2$ plasma for various structured electrodes. Solid lines are visual guidelines.

Figure 4 shows that the self-bias potential for the corrugated structured electrodes decreased to zero at higher pressure and increased to positive voltage as shown later. The matching network attached to the rf power supply could not read the positive bias voltage and it reads zero as the lowest voltage even when it is positive as confirmed by voltage data measured at the dc bias supply.

The increase in the rf power does not change the self-bias potential measurement in the matching network as shown in Fig. 5, where self-bias potential variation with rf power at 450 mTorr pressure in Ar/$Cl_2$ plasma is shown.

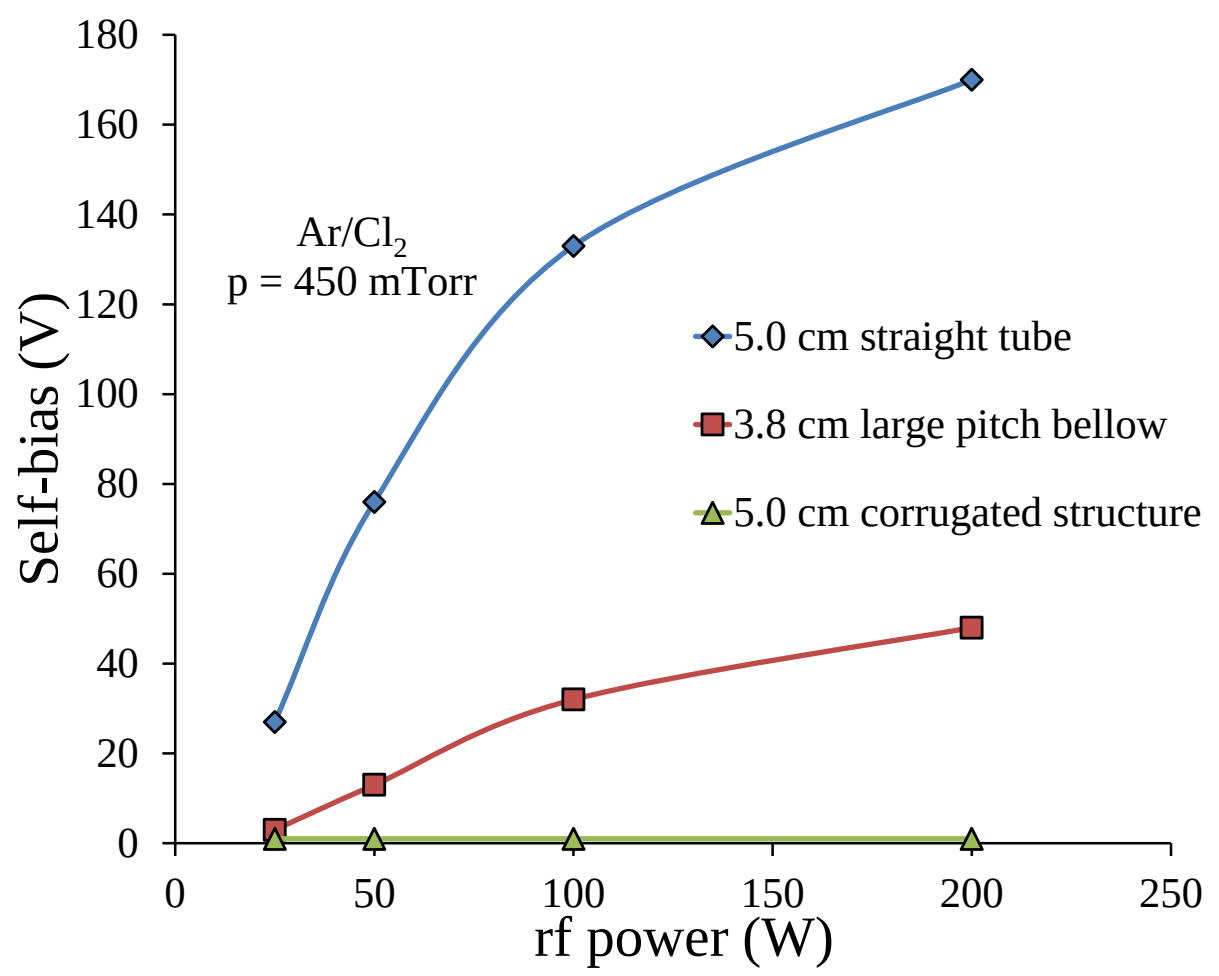

FIG. 5. Self-bias dependence in the $Ar/Cl_2$ plasma on the rf power for various structured electrodes. Solid lines are visual guidelines.

Figure 5 shows that there is no self-bias potential developed across corrugated structure for a high pressure $Ar/Cl_2$ plasma. It is to be noted that the increasing power does not change the situation, instead the bias potential turns positive and increases in the positive direction as shown in Fig. 7.

The self-bias potential can be brought to zero or a positive value by providing additional dc current to the inner electrodes. The dc current required to bring the self-bias potential to zero is plotted in Fig. 6 as a function of rf power at constant pressure of 450 mTorr.

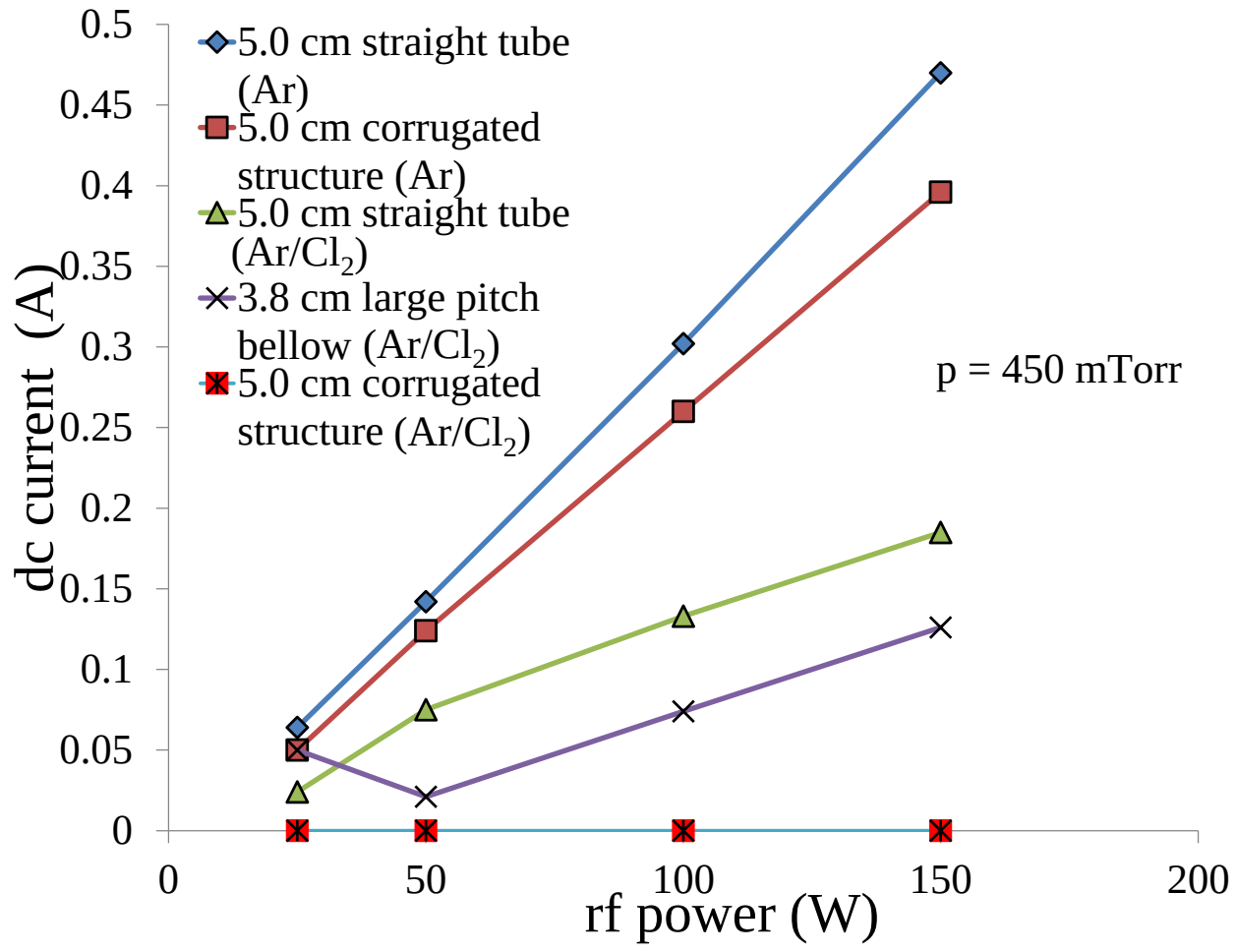

FIG. 6. The dc current variation with rf power for various structured electrodes and two different gas compositions. Solid lines are visual guidelines.

Figure 6 shows that though the current needed to make self-bias zero for Ar plasma is high compared to $Ar/Cl_2$ plasma due to its lower electron density, there is no current needed for $Ar/Cl_2$ plasma at higher pressure for the corrugated structure. The increase in rf power does not change this situation although increase in rf power show a dc voltage increase on dc power supply. The tube electrode and large pitch bellows electrode show increase in dc current requirement with the increase in rf power and show no change in dc voltage reading, which is zero. This positive dc voltage in the case of corrugated structure electrode can be described as a positive dc self-bias voltage. Considering this, we plotted the dc bias voltage for the structured electrodes in Fig. 7, where the dc self-bias voltage is plotted negative so that the reversal of the self-bias potential can be shown in the plot.

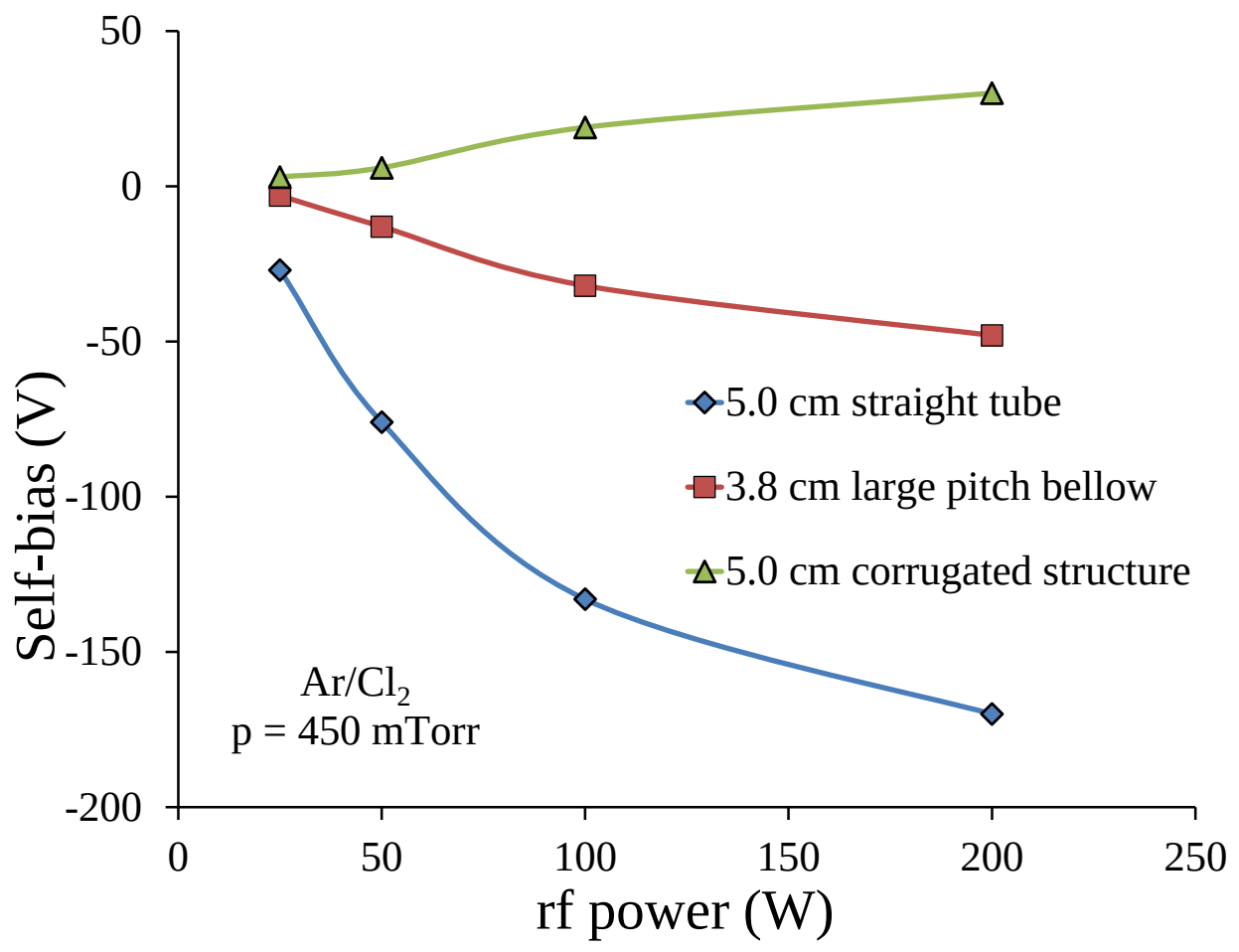

FIG. 7. Self-bias dependence on the rf power in the $Ar/Cl_2$ plasma for various structured electrodes. Solid lines are visual guidelines.

Figure 7 shows that, in $Ar/Cl_2$ plasma at higher pressure, the use of a corrugated structure electrode leads to the reversal of the asymmetry in the plasma. Instead of the negative dc self-bias potential, it acquires the positive dc self-bias potential, which increases with rf power.

Maintaining similar plasma conditions, the etch rate of Nb placed on the outer wall was measured while only the electrode shape was varied. The graph of the etch rate versus the different structure electrodes is shown in Fig. 8. The Nb etch rate increased fourfold, compared to the straight tube, when the corrugated structure was applied.

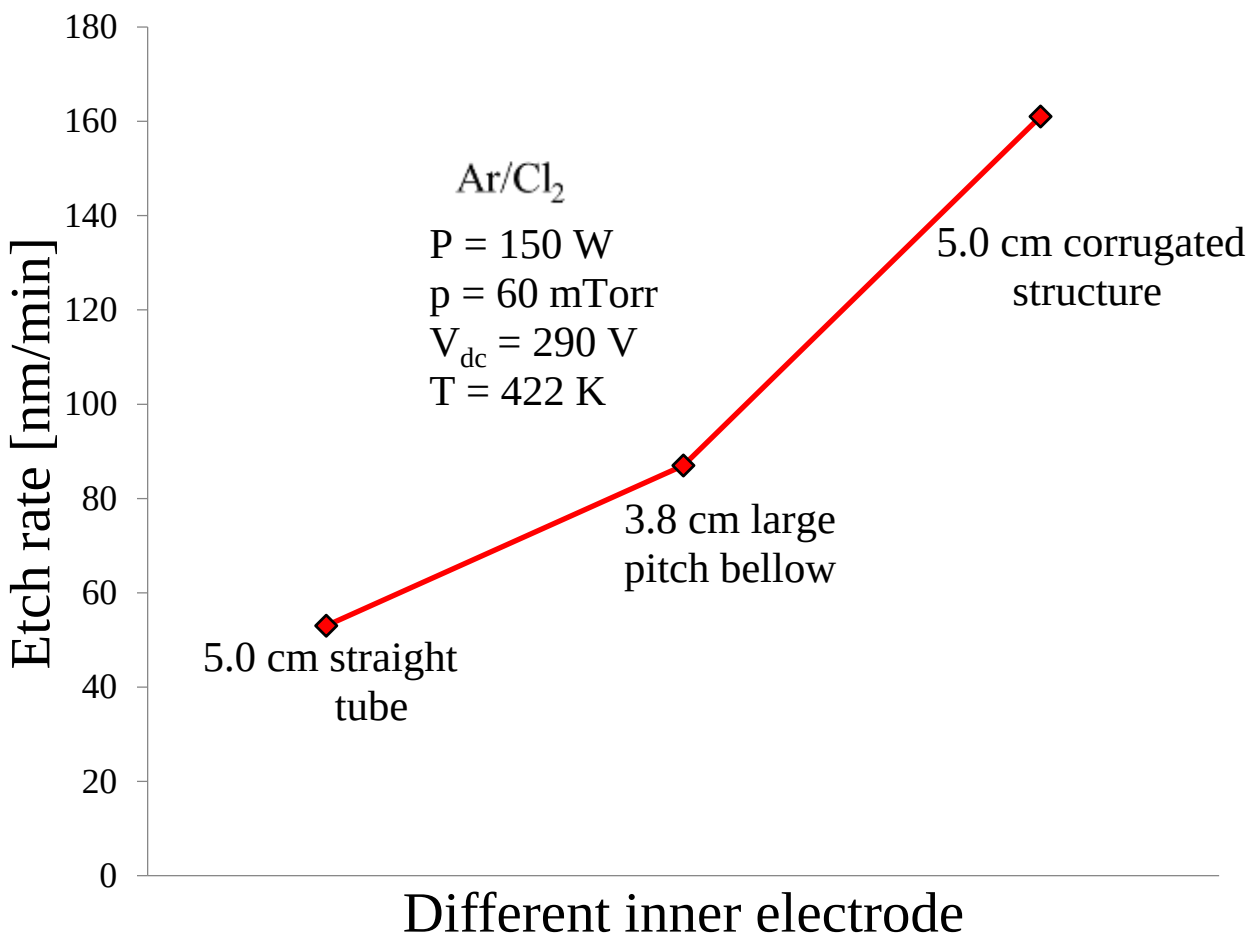

FIG. 8. The etch rate variation with various structured electrodes for the same plasma conditions. Solid line is a visual guideline.

In this study we explored the effect of geometrical enhancement of the surface area of the inner electrode and investigated the possibility to reduce the asymmetry in a coaxial CCP reactor. The geometrical modification of the inner electrode surface area to change the asymmetry of the coaxial cylindrical discharge is a novel concept. It is a cylindrical analogue to planar asymmetry corrections, which were studied in detail in Ref. [13]. The effect of different shaped inner electrode on the self-bias potential at different pressure and rf power is measured and the corrugated structure is found to be optimal for reducing and reversing the asymmetry. The addition of $Cl_2$ makes the asymmetry significantly smaller at higher pressure and rf power. It is shown that for the DLCS inner electrode there is no need to provide additional current to zero the self-bias potential. At higher pressure Ar/$Cl_2$ plasma, the dc self-bias potential, which is negative with a powered inner electrode, reacts to the reversal of the asymmetry in the plasma by becoming positive. The etch rate was measured for the same positive dc bias at lower pressure for various structured inner electrodes. The etch rate is higher for the DLCS electrode due to the reduction in plasma asymmetry.

*This work is supported by the Office of High Energy Physics, Office of Science, Department of Energy under Grant No. DE-SC0007879. Thomas Jefferson National Accelerator Facility, Accelerator Division supports J. Upadhyay through fellowship under JSA/DOE Contract No. DE-AC05-06OR23177.

[1] M. V. Alves, M. A. Lieberman, V. Vahedi, and C. K. Birdsall, J. Appl. Phys. 69, 3823 (1991).
[2] M. A. Lieberman and S. E. Savas, J. Vac. Sci. Technol. A 8, 1632 (1990).
[3] Y. P. Raizer and M. N. Shneider, Plasma Sources Sci. Technol. 1, 102 (1992).
[4] J. W. Coburn and E. Kay, J. Appl. Phys. 43, 4965 (1972).
[5] K. Kohler, J. W. Coburn, D. E. Horne, E. Kay, and J. H. Keller, J. Appl. Phys. 57, 59 (1985).
[6] H. M. Park, C. Garvin, D. S. Grimard, and J.W. Grizzle, J. Electrochem. Soc. 145, 4247 (1998).
[7] M. Zeuner, H. Neumann, and J. Meichsner, J. Appl. Phys. 81, 2985 (1997).
[8] J. Upadhyay, D. Im, S. Popović, A.-M. Valente-Feliciano, L. Phillips, and L Vusković, Phys. Rev. ST Accel. Beams 17, 122001 (2014).
[9] J. Upadhyay, D. Im, S. Popović, A.-M. Valente-Feliciano, L. Phillips, and L Vusković, J. Appl. Phys. 117, 113301 (2015).
[10] R. Hytry and D. BoutardGabillet, Appl. Phys. Lett. 69, 752 (1996).
[11] J. Upadhyay, D. Im, S. Popović, A.-M. Valente-Feliciano, L. Phillips, and L Vusković, arxiv: 1506.05167.
[12] E. V. Barnat, G. R. Laity and S. D. Baalrud, Physics of Plasmas 21, 103512 (2014).
[13] N. Schmidt, J. Schulze, E. Schungel and U. Czarnetzki, J. Phys. D: Appl. Phys. 46, 505202 (2013).
[14] J. C. Martz, D. W. Hess, and W. E. Anderson, J. Appl. Phys. 67, 3609 (1990).
[15] N. L. Bassett and D. J. Economou, J. Appl. Phys. 75, 1931 (1994).
[16] G. Franz, J. Vac. Sci. Technol. A 23, 369 (2005).
[17] M. V. Malyshev and V. M. Donnelly, J. Appl. Phys. 90, 3 (2001).